\documentclass[]{interact}

\usepackage{epstopdf}% To incorporate .pdf illustrations using PDFLaTeX, etc.
\usepackage[caption=false]{subfig}% Support for small, `sub' figures and tables
\usepackage[whole]{bxcjkjatype}
\usepackage{amsmath, amsfonts, amssymb}
\usepackage{multirow}
\usepackage{threeparttable}
\usepackage{color}

\usepackage{hyperref}
\usepackage[numbers,sort&compress]{natbib}% Citation support using natbib.sty
\bibpunct[, ]{[}{]}{,}{n}{,}{,}% Citation support using natbib.sty
\renewcommand\bibfont{\fontsize{10}{12}\selectfont}% Bibliography support using natbib.sty
\makeatletter% @ becomes a letter
\def\NAT@def@citea{\def\@citea{\NAT@separator}}% Suppress spaces between citations using natbib.sty
\makeatother% @ becomes a symbol again

\theoremstyle{plain}% Theorem-like structures provided by amsthm.sty

\theoremstyle{definition}

\theoremstyle{remark}

\begin{document}

\articletype{Regular article}
% Specify the article type or omit as appropriate

\title{Sparse mixed linear modeling with anchor-based guidance for high-entropy alloy discovery}

\author{
    \name{Ryo Murakami\textsuperscript{a}\thanks{CONTACT Ryo Murakami Email: MURAKAMI.Ryo@nims.go.jp}, Seiji Miura\textsuperscript{b}, Akihiro Endo\textsuperscript{a} and Satoshi Minamoto\textsuperscript{a}}
    \affil{
        \textsuperscript{a}Materials Data Platform, Research Network and Facility Services Division, National Institute for Materials Science, Tsukuba 305-0044, Ibaraki, Japan\\
        \textsuperscript{b}Division of Materials Science and Engineering, Faculty of Engineering, Hokkaido University, Sapporo 060-8628, Hokkaido, Japan
    }
}

\maketitle

\begin{abstract}
High-entropy alloys have attracted attention for their exceptional mechanical properties and thermal stability. However, the combinatorial explosion in the number of possible elemental compositions renders traditional trial-and-error experimental approaches highly inefficient for materials discovery. To solve this problem, machine learning techniques have been increasingly employed for property prediction and high-throughput screening. Nevertheless, highly accurate nonlinear models often suffer from a lack of interpretability, which is a major limitation. In this study, we focus on local data structures that emerge from the greedy search behavior inherent to experimental data acquisition. By introducing a linear and low-dimensional mixture regression model, we strike a balance between predictive performance and model interpretability. In addition, we develop an algorithm that simultaneously performs prediction and feature selection by considering multiple candidate descriptors. Through a case study on high-entropy alloys, this study introduces a method that combines anchor-guided clustering and sparse linear modeling to address biased data structures arising from greedy exploration in materials science.
% 近年、ハイエントロピー合金（High-Entropy Alloys, HEAs）は、優れた機械的特性や熱安定性から注目を集めている。しかし、構成元素の組み合わせ数が指数関数的に増加するため、従来の試行錯誤的な実験アプローチによる材料探索は非効率である。この課題に対処するため、機械学習を用いた物性予測やスクリーニング手法の研究が進められているが、高精度な非線形モデルは往々にして解釈性を欠くという問題がある。本研究では、材料実験データに内在する「貪欲的探索」に起因する局所的なデータ構造に着目し、線形かつ低次元の混合回帰モデルを導入することで、表現力と解釈性の両立を図る。また、複数の特徴量候補を用意し、予測と同時に重要な特徴量の選択を行うアルゴリズムを実施した。本研究は、HEAを対象とした具体例を通じて、探索バイアスを考慮した予測モデルの構築手法を示すとともに、解釈性と性能を兼ね備えた機械学習モデルの設計指針を提供するものである。
\end{abstract}


\section{Introduction}
In recent years, high-entropy alloys (HEAs) have garnered attention as next-generation materials for their outstanding mechanical properties, thermal stability, and corrosion resistance \cite{HEAreview2014, HEAreview2017}. Unlike conventional alloy designs, HEAs---also referred to as multi-principal element alloys---comprise multiple (typically five or more) principal elements, offering a high degree of chemical and structural freedom. This unique composition enables the exploration of novel properties unattainable in traditional materials systems.

However, the compositional design space of HEAs is vast, with the number of potential elemental combinations increasing exponentially with the number of constituent elements. For instance, the number of possible combinations of five elements selected from a pool of 60 metals is $_{60}C{_5} = 5,461,512$. Given the staggering scale, an exhaustive exploration of the space is impractical. Moreover, expanding the search to include non-equimolar compositions further increases the dimensionality of the space. As a result, trial-and-error based experimental exploration is highly inefficient and limited in its capacity for making new discoveries.

In addressing this issue, the application of machine learning techniques to the property prediction and feasibility assessment of HEAs has become an active area of research \cite{rao2022, Singh2023, kamnis2025}. A variety of non-linear and high-dimensional models---including deep learning, gradient boosting, and kernel based methods---have achieved high prediction accuracy. Despite their performance, these complex models often suffer limited interpretability due to their black-box nature.

Conversely, linear and low-dimensional models offer high interpretability but frequently fall short in terms of predictive performance, particularly in practical scenarios involving materials discovery \cite{ward2022, muckley2023}. One contributing factor is the biased nature of data collection in materials science, which often follows from a greedy exploration strategy. In adopting this strategy, once a promising material is identified, compositions in its vicinity are preferentially investigated, leading to the development of numerous variations of similar materials \cite{thompson2023}. This results in a locally biased dataset, making it difficult to represent the global structure of the compositional space with a single linear model. Consequently, non-linear and high-capacity models are commonly employed to accommodate such data biases.

For example, the Cantor alloy \cite{cantor2004}, a prototypical HEA, has attracted attention for its ductility, toughness, and strength at low temperatures, prompting extensive investigations into related alloys---such as those involving Ni substitution or Cr content adjustments. Similarly, the Senkov alloy system \cite{senkov2010, Senkov2011} has been explored extensively for its superior high-temperature strength, leading to systematic studies on refractory metal based compositions. These examples highlight how initial discoveries drive focused exploration in their compositional neighborhoods, resulting in clustered data distributions within limited regions of the vast design space. This inherent data bias implies that collected datasets may not adequately represent the entire search space, thus necessitating non-linear, high-dimensional modeling approaches.

Given this context, it is crucial to develop predictive models that are both interpretable and adaptable to the biased structure of data generated through greedy exploration. Maintaining the interpretability of linear models while flexibly addressing locally biased data structures requires the capture of region-specific correlations within the compositional space. A promising approach in this regard is the use of mixture regression models \cite{mcLachlan2000, fraley2002, douglas2015, hirakawa2021}, where distinct linear regressions are applied to different regions of the search space. By partitioning the space into local domains and fitting appropriate linear models to the different domains, mixture regression captures global non-linearities while preserving the interpretability of the components.

In this study, we investigate the structural characteristics of datasets generated through greedy exploration and introduce a mixture regression framework tailored to this context. Our goal is to balance predictive performance and interpretability while flexibly adapting to the underlying structure of real-world exploration data.

Furthermore, in the field of materials science, understanding why certain properties emerge is as important as accurate prediction. Understanding the emergence of properties is essential for establishing design principles and guiding the development of novel materials. However, the key features that correlate with HEA properties are often not obvious. To ensure physical plausibility and enable feedback into materials design, it is vital to build models using a small number of physically meaningful features. Accordingly, we integrate feature selection within the mixture linear modeling framework \cite{nagata2015, obinata2022}, enabling both interpretability and adaptability to structurally biased data.

In summary, this work proposes a feature-selective mixture linear model that accommodates the biased structure of HEA datasets resulting from greedy exploration. Through validation with real-world data, we demonstrate the potential of our approach to simultaneously achieve interpretability and predictive performance. Our approach thus offers a practical and effective framework for supporting materials design under biased data conditions.

\begin{figure}
    \centering
    \includegraphics[width=\linewidth]{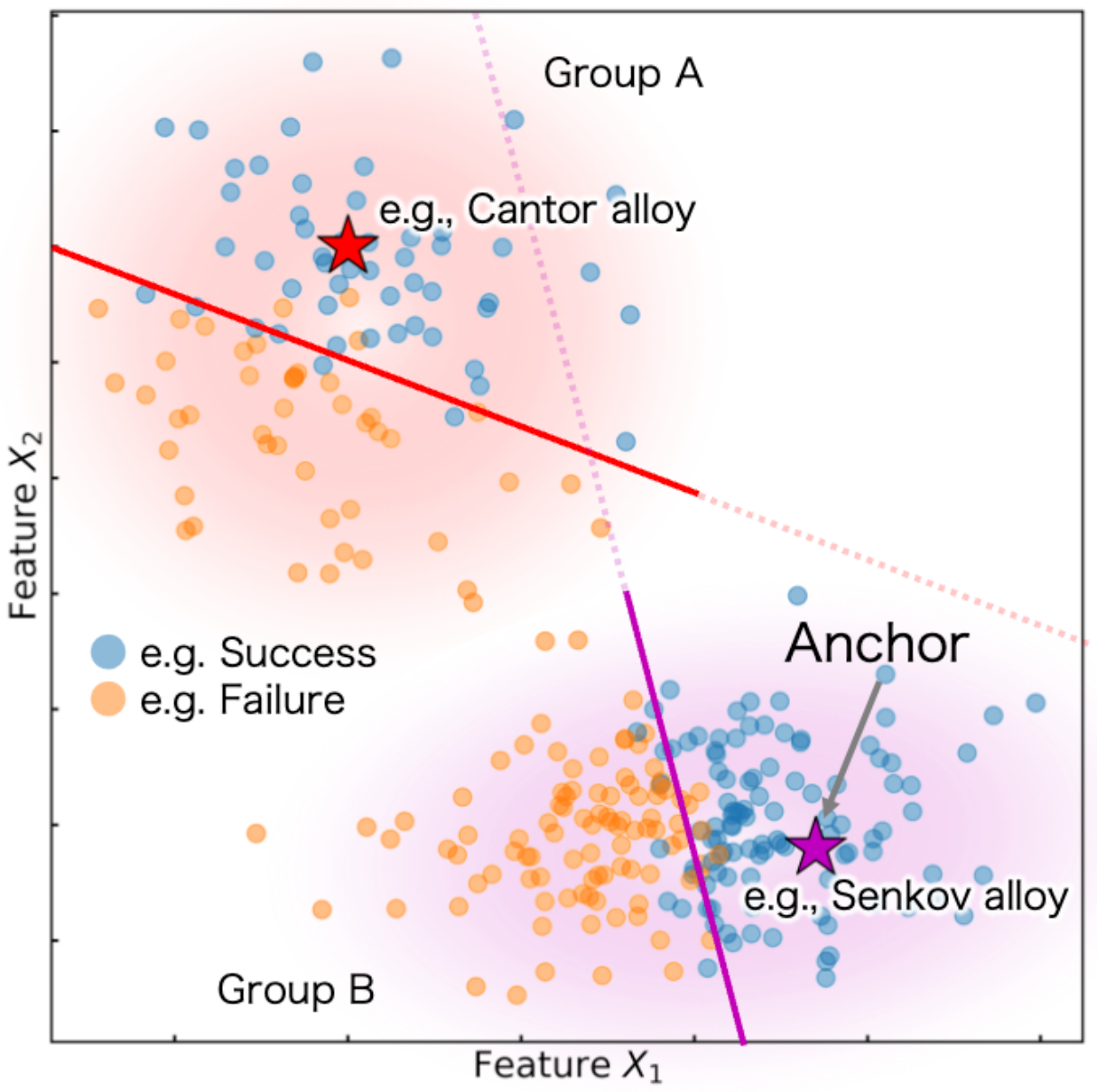}
    \caption{Conceptual diagram of the mixed model referring to anchor data used in this study. The task addresses the problem of binary classification. Note that this method can be used not only for a classification task but also for a regression task. }
    \label{fig:strategy}
\end{figure}

\section{Research strategy}
In many cases, experimental datasets for materials development are collected adopting greedy exploration strategies. In this study, we investigate the structural characteristics of datasets generated by greedy exploration and introduce mixture linear modeling in which each cluster is represented by a linear model with distinct parameter values. An overview of the proposed modeling strategy is illustrated in Figure \ref{fig:strategy}. As shown, multiple linear models are locally applied across different regions of the feature space. This enables the modeling to flexibly adapt to the actual structure of exploration data while balancing expressiveness and interpretability.

However, a key challenge in mixture linear modeling lies in defining clusters, which is often ambiguous and difficult to interpret in a physically meaningful way. Traditional approaches sequentially estimate cluster labels for each data point as part of the model. However, this introduces a number of problems: the number of variables increases with the dataset size, the optimization landscape becomes highly non-convex with many local minima, and the resulting clusters may lack physical significance, thereby limiting interpretability.

To address these issues, we turn our attention to a characteristic pattern frequently observed in materials exploration: greedy search behavior. Specifically, when a material with promising properties is discovered, compositions or features in its vicinity tend to be intensively explored. Therefore, we identify representative materials implicitly selected by researchers as exploration anchors (denoted by stars in Figure \ref{fig:strategy}). We then perform distance-based clustering in the feature space centered around these anchor points. By fixing these physically grounded clusters, we construct a mixture linear model that provides improved interpretability and aligns with the underlying data generation process. This approach enables us to proactively leverage the biased structure induced by greedy exploration, thereby bridging the gap between clustering uncertainty and model interpretability.

Nevertheless, it is not immediately obvious which subset of features is most relevant for both accurate prediction and meaningful clustering in the context of HEAs. In high-dimensional spaces, models are prone to overfitting, which undermines generalization performance. In addition, data points tend to be sparsely distributed, making it difficult to identify similar samples. To address this, we consider a wide set of candidate features and aim to simultaneously perform both prediction and feature selection. In this study, we employ a Bayesian sparse modeling framework for feature selection, allowing us to identify a compact set of physically meaningful descriptors that contribute to both accurate prediction and interpretable clustering.

% 多くの材料開発の実験データセットは，貪欲的な探索により集められることが多い．そこで本研究では，「貪欲的探索に基づくデータセット構造の特性」に着目すし，クラスターごとにパラメータ値の異なったモデルで表現する混合線形モデルを導入する．図\ref{fig:strategy}に混合線形モデルの概要図を示す．図に示すように，複数の線形モデルを局所的に貼り付けることで，モデルの表現と解釈性の両立を図りつつ，実際の探索データの構造に適応した柔軟なモデル構築を実現する．

% 一方で，混合線形モデルにおける「クラスター」の定義は，モデル設計においてしばしば曖昧であり，明確な物理的解釈を持たせることが難しい．従来の手法では，各データ点のクラスタラベルをモデルの一部として逐次的に推定するが，このアプローチはデータ点数に比例して変数が増加し，局所解が多発するという問題を引き起こす．さらに，推定されたクラスターが物理的な意味を有するとは限らず，解釈性の観点からも課題が残る．本研究では，このような課題に対し，材料探索にしばしば見られる「貪欲的探索パターン」に着目する．すなわち，良好な性能を示す材料が見出された後，その周辺の化学組成や特徴量空間が重点的に探索されるという傾向である．この観察に基づき，我々は研究者が探索の「アンカー」（図\ref{fig:strategy}における星記号）として選択した代表的な材料群を中心に，特徴量空間上で距離に基づくクラスタリングを実施する．その結果として得られたクラスタを固定し，混合線形モデルを構築することで，物理的な意味づけが可能な解釈性の高いモデルを実現する．この手法により，貪欲的探索によって形成されたデータ構造を積極的に活用し，クラスタリングの不確実性と解釈性のギャップを埋めることを目指す．

% 一方で，どのような特徴量空間（特徴量サブセット）がHEAの予測とクラスター形成に有効であるかは自明ではない．高次元空間では，モデルが学習データにオーバーフィットしてしまい汎化性能が低下したり，データ点同士が非常に離れてしまうため似たようなデータが見つけにくくなる．そこで我々は，特徴量候補を複数個用意し，HEAの予測だけでなく重要な特徴量の選択も同時に行う．本研究ではベイズ推論によるスパースモデリングにより，特徴量選択を実行する．

\section{Analysis setup}

\subsection{Dataset and task---Single- or multi-phase?}
The prediction of phase---specifically, the prediction of whether a given composition forms a single-phase or multi-phase structure---is a central task in the design of HEAs. Our task is to estimate the phase label $y_n \in \{0, 1\}$ ($y_n=1$ for single-phase, $y_n=0$ for multi-phase) using a feature set $X_n \in \mathbb{R}^{M}$ derived from composition information. The dataset is defined as $\mathcal{D} = \{(y_n, X_n)\}_{n=1}^{N}$, where $N$ is the number of data. This dataset was obtained from literature \cite{Singh2023} and is outlined in Appendix \ref{appx:dataset}.

The analytical model used for prediction is the Bayesian logistic regression model expressed as
\begin{align}
    y_n &\sim \mathcal{B}(y_n \mid p_n) = p_{n}^{y_n}(1 - p_{n})^{1-y_n}, \\
    p_n &= p(X_{n}; \Theta) = \frac{1}{1+\exp{ \{ - f(X_{n}; \Theta) \} }},
\end{align}
where $\Theta$ is the parameter set. The probability $\mathcal{B}(y_n \mid p_n)$ follows a Bernoulli distribution. $f(X_{n}; \Theta)$ is the analytical model.

In this study, we randomly divided the 1200 data into 1080 training data (90\% of the 1200 samples) and 120 test data (10\% of the 1200 samples) to construct a prediction model. Table \ref{table:feature_set} gives the content of the feature set and descriptions. We used 15 features with physical meaning as explanatory variables. Moreover, Table \ref{table:anchor_set} presents the anchor data used in this study.

\begin{table}
    \centering
    \caption{Content of the feature set (10 features) used in this study and descriptions}
    \label{table:feature_set}
    \scalebox{1.0}[1.0]{
    \begin{tabular}{c|cc}
        \hline
        Symbol & Source & Description \\
        \hline
        $ \Delta H_{\mathrm{mix}} $               & \multirow{5}{*}{Paper \cite{Singh2023}} & Mixing enthalpy\\
        $ \Delta S_{\mathrm{mix}} $               & & Mixing entropy \\
        $ \delta                  $               & & Atomic size difference \\
        $ \Delta \chi_{\mathrm{Pauling}} $        & & Pauling's electronegativity difference \\
        $ VEC $                                   & & Valence electron concentration \\
        \hline
        $ \langle \chi_{\mathrm{Allen}} \rangle$  & \multirow{10}{*}{XenonPy \cite{XenonPy2019}} & Allen's electronegativity mean \\
        $ \Delta \chi_{\mathrm{Allen}}$           & & Allen's electronegativity difference \\
        $ \langle N_{\mathrm{Valence}} \rangle$   & & Total valence electron mean \\
        $ \Delta N_{\mathrm{Valence}}$            & & Total valence electron difference \\
        $ \langle P_{\mathrm{Periodic}} \rangle$  & & Period mean (in the periodic table) \\
        $ \Delta P_{\mathrm{Periodic}}$           & & Period difference (in the periodic table) \\
        $ \langle T_{\mathrm{Melting}} \rangle$   & & Melting point mean \\
        $ \Delta T_{\mathrm{Melting}}$            & & Melting point difference \\
        $ \langle N_{\mathrm{Mendeleev}} \rangle$ & & Mendeleev number mean \\
        $ \Delta N_{\mathrm{Mendeleev}}$          & & Mendeleev number difference \\
        \hline
    \end{tabular}
    }
    \begin{tablenotes}
      \footnotesize
      \item \hfill Note: $\langle x \rangle$ and $\Delta x$ are the weighted mean and the weighted variance based on the molar ratio.
    \end{tablenotes}
\end{table}

\begin{table}
    \centering
    \caption{Anchor data (two samples) used in this study and descriptions. }
    \label{table:anchor_set}
    \scalebox{1.0}[1.0]{
    \begin{tabular}{c|ccc}
        \hline
        Anchor & Alloy       & Crystal   & Phase \\
        name   & composition & structure & structure \\
        \hline
        Senkov alloy \cite{senkov2010} & $\mathrm{Hf_{0.2}Nb_{0.2}Ta_{0.2}Ti_{0.2}Zr_{0.2}}$ & BCC & Single phase \\
        Cantor alloy \cite{cantor2004} & $\mathrm{Co_{0.2}Cr_{0.2}Fe_{0.2}Mn_{0.2}Ni_{0.2}}$ & FCC & Single phase \\
        \hline
    \end{tabular}
    }
    \begin{tablenotes}
      \footnotesize
      \item \hfill Note: BCC and FCC refer to body-centered cubic and face-centered cubic crystal structures, respectively.
    \end{tablenotes}
\end{table}

\section{Method}

\subsection{Model---Mixed linear model referring to anchors}
We describe an analysis model $f_{\bm{g}}(X_{n}; \Theta)$, which is called the mixed linear model. The model $f_{\bm{g}}(X_{n}; \Theta)$ is expressed as
\begin{align}\label{MLM}
    f_{\bm{g}}(X_{n}; \Theta) = \sum_{k=1}^{K}{ g_{n,k} \left( w_k^{\top}X_{n} \right) } + b,
\end{align}
where the parameter set $\Theta$ is $\{\{w_1,w_2,\cdots,w_K\}, b\}$, $w_k \in \mathbb{R}^{M}$ denotes the weight coefficients, $b \in \mathbb{R}$ is the bias term, the matrix $\bm{g} \in \{0, 1\}^{N \times K}$ is a binary matrix assigning each data point to a group, and $K$ is the number of groups.

In our modeling approach, the group structure (represented by the binary matrix $\bm{g}$) is predefined prior to parameter estimation. The analyst specifies representative data points, referred to as anchor data  $A_j$, for each group. Clustering is then carried out according to the distance to these anchors, with each data point assigned to the nearest anchor-defined group. The element $g_{n, k}$ of the binary matrix $\bm{g}$ is expressed as
\begin{equation}
    g_{n, k} = 
    \begin{cases}
        \: 1 & \text{if} \:\:\: k = \underset{j \in [K]} {\operatorname{argmin}} \: D(X_n, A_j) \\
        \: 0 & \text{otherwise},
    \end{cases}
\end{equation}
where $[K]$ denotes $\{1, 2, \cdots, K\}$ and the function $D(X_n, A_j)$ is the distance function for calculating the distance between vectors $X_n$ and $A_j$. In this study, we employed the Euclidean distance $D(X_n, A_k) = \|X_n - A_k\|_{2}$ where $\|\cdot\|_{2}$ denote $L^{2}$-norm. If $g_{n,k}=1$, the $n$-th data point belongs to the $k$-th anchor group. Here, $\sum_{[K]}{g_{n, k}}$ = 1. $K$ is the number of anchors.

\subsection{Algorithm---How to select a model and feature set}
We select the feature subset $\mathcal{S} \subseteq \{1, 2, \cdots, M\}$ from the hypothesis set $\mathcal{H} = \{\mathcal{S} \mid 1 \leq |\mathcal{S}| \leq L\}$ \cite{igarashi2018}. We assume that we can describe the data using at most $L$ features and perform an exhaustive search. We have to design the evaluation value of the feature subset $\mathcal{S}$ to select the feature subset $\mathcal{S}$. This study used the Bayesian free energy $F_{\mathcal{S}} = - \ln{ P(\mathcal{S} \mid \mathcal{D}) }$ (the negative logarithm marginal posterior probability) in Bayesian estimation. The Bayesian free energy $F_{\mathcal{S}}$ of the feature subset $\mathcal{S}$ is expressed as
\begin{align}
    F_{\mathcal{S}} &= - \ln{ P(\mathcal{S} \mid \mathcal{D}) }, \\
                    &= - \ln{ \int{P(\Theta, \mathcal{S} \mid \mathcal{D})}\mathrm{d}\Theta }, \\
                    &\varpropto - \ln{\int{P(\mathcal{D} \mid \Theta, \mathcal{S})P(\Theta)P(\mathcal{S})}\mathrm{d}\Theta },
\end{align}
where $P(\Theta, \mathcal{S} \mid \mathcal{D})$ and $P(\mathcal{D} \mid \Theta, \mathcal{S})$ denote the posterior distribution and likelihood under the model $f_{\bm{g}}(X_{n}^{(\mathcal{S})}; \Theta)$. Here, $X_{n}^{(\mathcal{S})} \in \mathbb{R}^{|\mathcal{S}|}$ is a feature subvector of $X_{n}$. The probability distributions $P(\Theta)$ and $P(\mathcal{S})$ are the prior distributions. $P(\mathcal{S})$ is assumed to be a uniform distribution. Appendix \ref{appx:selection} presents the settings of the prior distribution $P(\Theta)$.

As exact integration over the parameter space is computationally challenging, we used the Watanabe Bayesian information criterion (WBIC) \cite{wbic2013} to calculate the Bayesian free energy approximately
\begin{align}
    F_{\mathcal{S}} &\simeq \int{ \mathcal{L}(\Theta)\left[\frac{P(\mathcal{D} \mid \Theta, \mathcal{S})^{\beta}P(\Theta)}{\int{ P(\mathcal{D} \mid \Theta, \mathcal{S})^{\beta}P(\Theta)\mathrm{d}\Theta}}\right] \mathrm{d}\Theta }, \\
                    &= \mathbb{E}_{P(\mathcal{D} \mid \Theta, \mathcal{S})^{\beta}P(\Theta)}[\mathcal{L}(\Theta)],
\end{align}
where $\mathcal{L}(\Theta)$ is the log-likelihood with a parameter set $\Theta$, $\beta$ is the inverse temperature introduced mathematically, and $\beta^{-1} = \ln{N}$. We can calculate the expected value $\mathbb{E}[\mathcal{L}(\Theta)]$ from the sampling result at $\beta^{-1} = \ln{N}$ obtained using a sampling method, such as the Hamiltonian Monte Carlo method \cite{Duane1987,Neal2011} or the Gibbs sampling method. We calculate the Bayesian free energy $F_{\mathcal{S}}$ for all states in the set of hypotheses $\mathcal{H}$. Furthermore, the marginal probability $P(\mathcal{S}_{m} \mid \mathcal{D})$, which captures the relevance of feature $X_{m}$, is expressed as
\begin{equation}
    P(\mathcal{S}_{m} \mid \mathcal{D}) = \sum_{\mathcal{S} \subseteq \mathcal{H}_{m}}{P(\mathcal{S} \mid \mathcal{D})},
\end{equation}
where $\mathcal{H}_{m} \subseteq{\mathcal{H}}$ is a hypothesis set that includes a feature $X_{m}$.

We also address the problem of model selection, specifically by comparing the performance of a simple linear model and a mixed linear model. The evaluation value of the model is expressed as
\begin{align}
    F_{\mathrm{model}} = - \ln{ P_{\mathrm{model}} } = - \ln{ \sum_{\mathcal{S} \in \mathcal{H}}{P(\mathcal{S} \mid \mathcal{D})} },
\end{align}
where $P(\mathcal{S} \mid \mathcal{D})$ is calculated from $F_{\mathcal{S}}$.

\section{Experimental configuration}
In this study, we used the Hamiltonian Monte Carlo sampling method \cite{Duane1987,Neal2011} to estimate the parameters and the WBIC. The Hamiltonian Monte Carlo method was implemented using NumPyro \cite{NumPyro2019}. The burn-in period was set to 5000 iterations, and 1000 samples were drawn for inference. The computations were performed on a MacBook Air equipped with an Apple M2 chip and 24 GB of RAM. Appendix \ref{appx:selection} gives the settings for the prior distribution.

\section{Results and discussion}

\begin{figure}
    \centering
    \includegraphics[width=\linewidth]{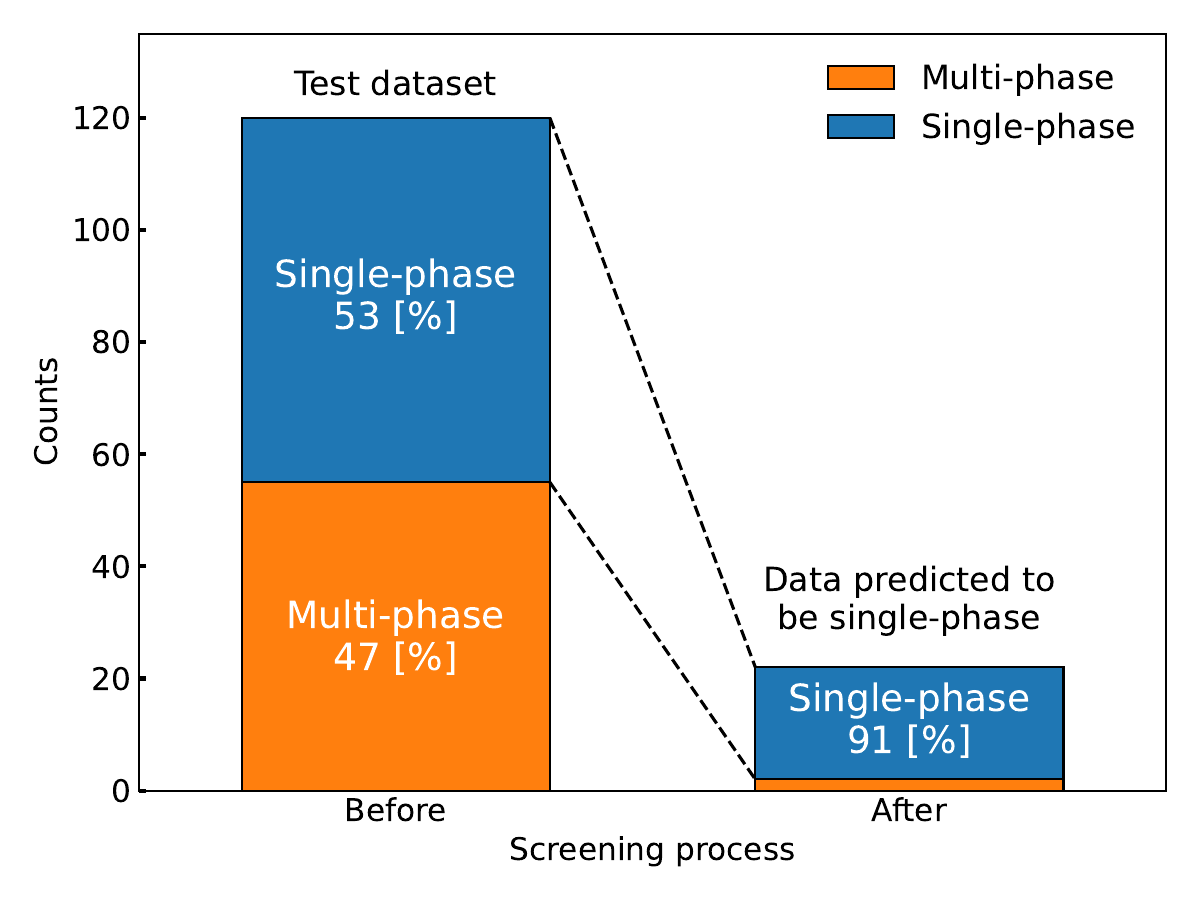}
    \caption{Results of applying the learned predictive model to the test data for screening. }
    \label{fig:screening}
\end{figure}

\subsection{Screening using a learning model}
We first demonstrate the screening performance of the trained predictive model. Figure \ref{fig:screening} presents the screening results obtained on the test dataset using the trained model. The threshold for the prediction model was set to 0.75. The x-axis represents the state before and after the screening process, whereas the y-axis indicates the number of samples in the dataset.

The "Before" condition refers to the original distribution of single-phase and multi-phase samples in the test data, whereas the "After" condition shows the phase distribution among the samples predicted as single-phase by the model. In the figure, prior to screening, the dataset comprises an approximately equal ratio of single-phase and multi-phase samples. After screening, however, the proportion of single-phase samples increases to approximately 90\%, indicating that the model enriches the concentration of single-phase candidates.

Although the model may overlook some true single-phase samples, the results reveal screening effectiveness sufficient for guiding the next stage of experimental planning. Representative performance metrics are provided in Appendix \ref{appx:metrics}.
% まず学習した予測モデルのスクリーニング性能を示す．本研究においては，バーンインは5000回，サンプリングは1000回に設定した．Figure \ref{fig:screening}に，テストデータに対して，学習した予測モデルを用いたスクリーニングした結果を示す．この図において，x軸はスクリーニング処理の前後を示し，y軸はデータのサンプル数を示す．スクリーニング処理のbeforeはテストデータの単相と複相の割合を示し，afterは予測モデルが”単相”と推定したデータセットの単相と複相の割合を示している．図に示すように，スクリーニング前のデータセットでは単相と複相の割合が約半分半分である．一方で，スクリーニング後のデータセットは単相の割合が9割となり，単相の濃度が非常に向上していることが確認できる．このモデルにより単相となるデータを見落としている可能性があるが，次時点の実験計画を進める上で，十分なスクリーニング効果を示している．Appendix \ref{appx:metrics}に代表的なメトリックを値を示す．

\begin{table}
    \centering
    \caption{Bayesian free energy $F_{\mathrm{model}}$ and marginalization probability $P_{\mathrm{model}}$ [\%] for the simple model (adopting only Bayesian logistic regression) and the mixed model (adopting our method). The bias term $b$ is included in the parameter set for the two models.}
    \label{table:model_selection}
    % \scalebox{1.1}[1.1]{
    \begin{tabular}{ccc|cc}
        \hline
        \multirow{2}{*}{Eq.} & Model & Parameter & Free energy & Probability \\
        & $f_{\bm{g}}(X_{n}; \Theta)$ & $\bm{w}$ & $F_{\mathrm{model}}$ & $P_{\mathrm{model}}$ [\%] \\
        \hline
        (\ref{eq:fixed}) & simple & $\bm{w} \in \mathbb{R}^{|\mathcal{S}|}$ & 737.72      & 0.00 \\
        (\ref{eq:random}) & mixed & $\bm{w} \in \mathbb{R}^{|\mathcal{S}|\times K}$ & \bf{685.82} & \bf{100.00} \\
        \hline
    \end{tabular}
    % }
    % \begin{tablenotes}
    %   \footnotesize
    %   \item \hfill*Our method
    % \end{tablenotes}
\end{table}

\subsection{Model selection using the marginal probability}
In this section, we compare our proposed method with Bayesian logistic regression in the context of HEAs, focusing on marginal probabilities. Table \ref{table:model_selection} presents the Bayesian free energies and marginal probabilities for both our method and Bayesian logistic regression.

Table \ref{table:model_selection} shows that our method outperforms conventional Bayesian logistic regression, achieving 100\% probability. It is noted that the marginal probabilities were obtained by integrating over both the model parameters and the subsets of features. The marginalization over the model parameters was performed using the WBIC. Details of the design of prior distributions and the results of model selection under different prior specifications are provided in Appendix \ref{appx:selection}.
% 本節では，ハイエントロピー合金において，我々の方法とベイズロジスティック回帰を周辺化確率に基づき比較する．Table \ref{table:model_selection}に，我々の方法とベイズロジスティック回帰におけるベイズ自由エネルギーおよび周辺化確率を示す．表より，通常のベイズロジスティック回帰より，我々の方法が100\%の確率で良好であることが分かる．ここで，周辺化確率はパラメータおよび特徴量サブセットに対して周辺化操作を実施した結果であることに注意されたい．また，パラメータに関する周辺化操作はWBICを用いて行いました．Appendix \ref{appx:selection}に事前分布の設計やその設計に違いに対するモデル選択結果を示す．

\begin{figure}
    \centering
    \includegraphics[width=\linewidth]{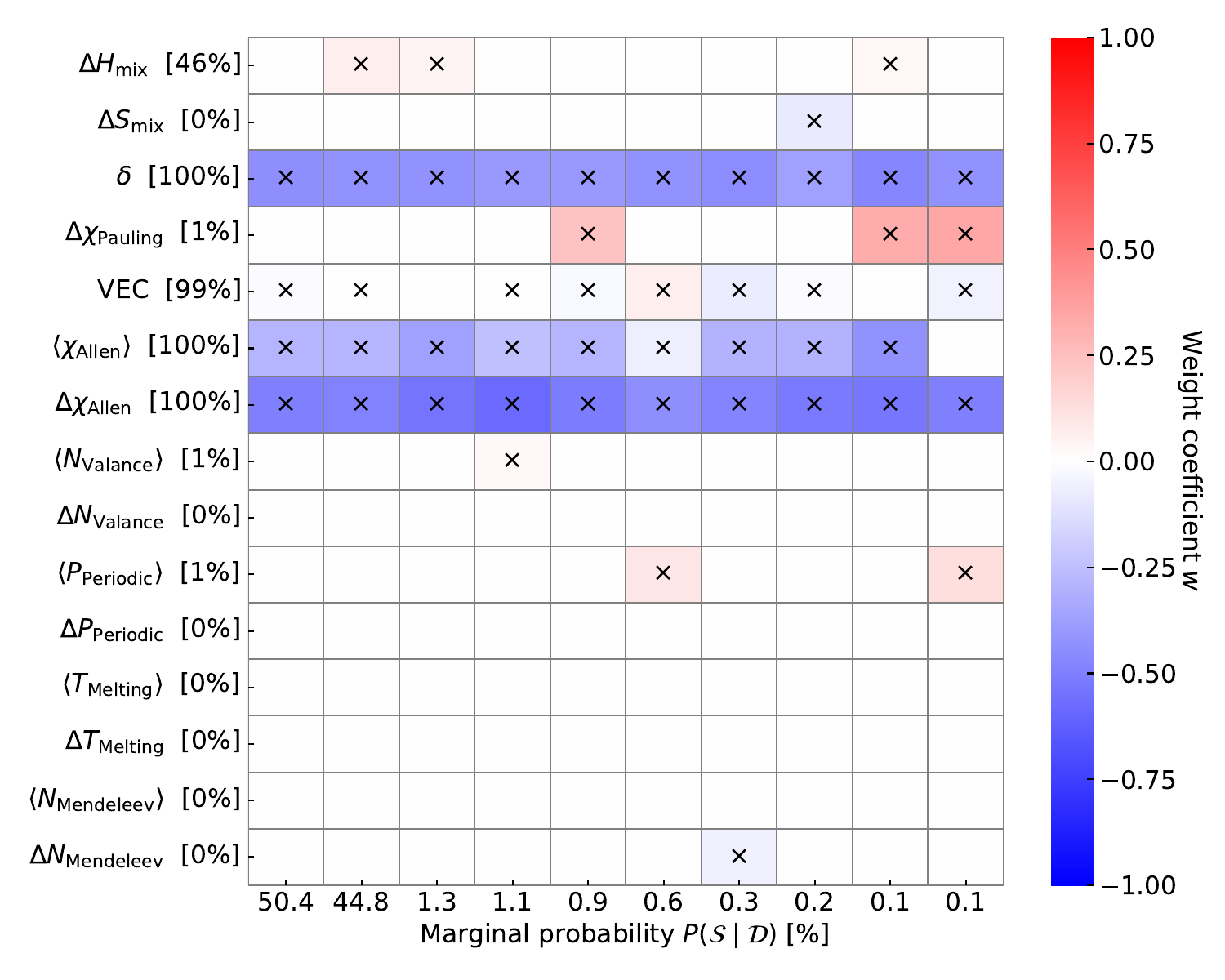}
    \caption{Feature selection using Bayesian inference through exhaustive searching, with the constraint of selecting no more than $L$ features. The value in square brackets [X\%] next to each feature label denote the marginalized probability of the feature being included in any subset; i.e., $P(\mathcal{S}_m \mid \mathcal{D}) = \sum_{\mathcal{H}_m} P(\mathcal{S} \mid \mathcal{D})$, where $\mathcal{H}_m \subseteq \mathcal{H}$ is the set of all feature subsets that include feature $m$. A cross mark on a cell indicates that the corresponding feature is included in a particular subset.}
    \label{fig:heatmap_mixture}
\end{figure}

\begin{figure}
    \centering
    \includegraphics[width=\linewidth]{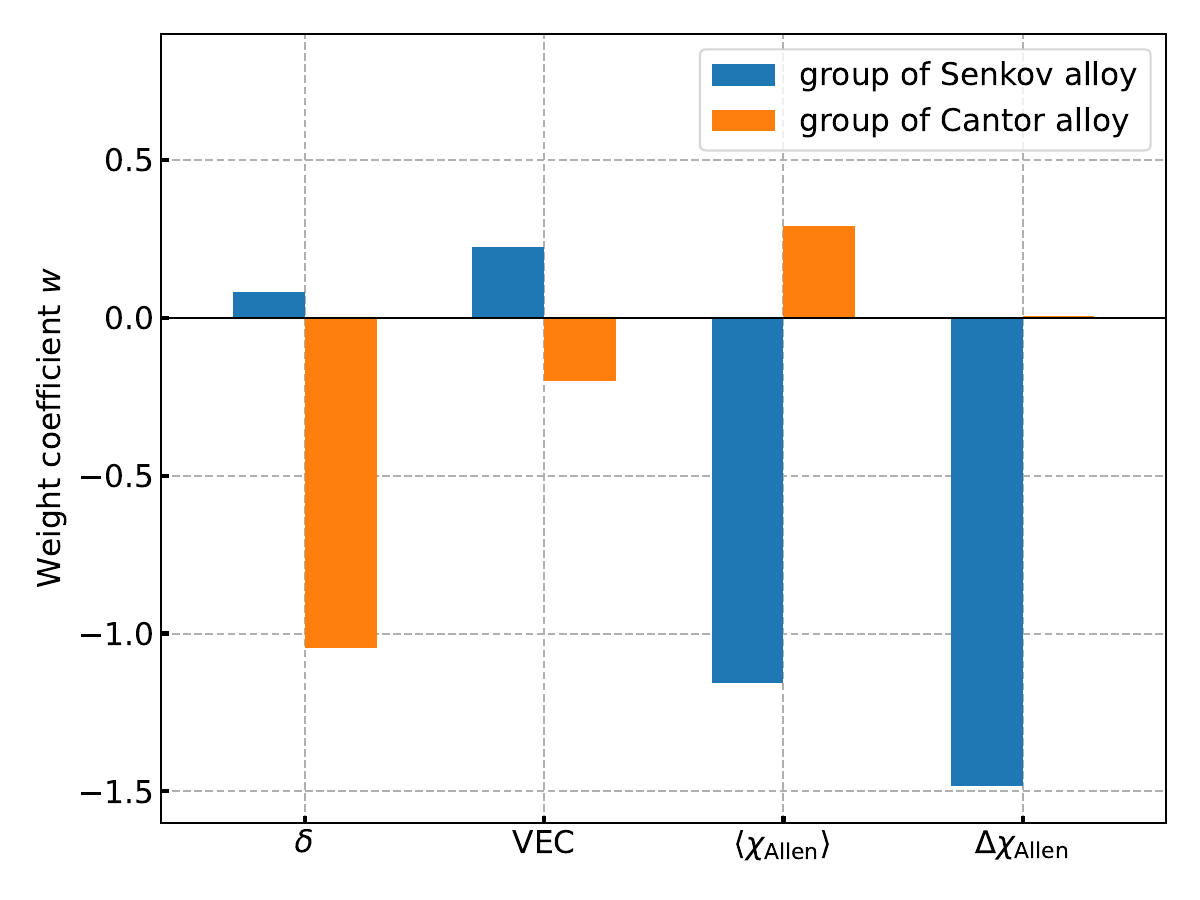}
    \caption{Weight coefficient parameter $w$ for the optimal feature subset $\{ \delta, VEC, \langle \chi_{\mathrm{Allen}} \rangle, \Delta\chi_{\mathrm{Allen}} \}$. }
    \label{fig:coef}
\end{figure}

\subsection{Feature selection and coefficient comparison}
In the proposed method, we perform a search over feature subsets during the training of the predictive model. In this study, the maximum number of features allowed in a subset was set to $L = 5$. The results of feature selection using Bayesian inference are shown in Figure \ref{fig:heatmap_mixture}. The x-axis represents the marginal probability $P(\mathcal{S} \mid \mathcal{D})$ obtained via model parameter marginalization, the y-axis indicates the feature labels, and the bar color (blue to red) corresponds to the weight parameter $w$. The weights $w$ are averaged across groups for visualization. Red and blue indicate positive and negative correlations, respectively. A cross mark on a cell indicates that the corresponding feature is included in a particular subset. The numbers in square brackets [X\%] next to each feature label denote the marginalized probability of the feature being included in any subset; i.e., $P(\mathcal{S}_m \mid \mathcal{D}) = \sum_{\mathcal{H}_m} P(\mathcal{S} \mid \mathcal{D})$, where $\mathcal{H}_m \subseteq \mathcal{H}$ is the set of all feature subsets that include feature $m$.

Figure \ref{fig:heatmap_mixture} shows that two feature subsets together account for approximately 95\% of the total probability (50.4\% + 44.8\%). The common features in these subsets are $\{ \delta, \mathrm{VEC}, \langle \chi_{\mathrm{Allen}} \rangle, \Delta\chi_{\mathrm{Allen}} \}$. The full set of candidate features includes pairs of similar features that convey analogous information but are defined differently, such as $\Delta\chi_{\mathrm{Allen}}$ and $\Delta\chi_{\mathrm{Pauling}}$, or VEC and $\Delta N_{\mathrm{Valence}}$. These types of features, referred to as similar features, are often removed prior to analysis in practice to avoid multicollinearity. However, it is typically unclear which definition is more appropriate, making this a nontrivial decision. In this study, as we exhaustively explored all possible combinations of feature subsets, we were able to assess the relevance of similar features directly. Figure \ref{fig:heatmap_mixture} shows that $\Delta\chi_{\mathrm{Allen}}$ and $VEC$ were particularly important in the prediction of HEAs compared to $\Delta\chi_{\mathrm{Pauling}}$ and $\Delta N_{\mathrm{Valence}}$.

Figure \ref{fig:coef} presents the weight parameters $w$ associated with the optimal feature subset $\{ \delta, \mathrm{VEC}, \langle \chi_{\mathrm{Allen}} \rangle, \Delta\chi_{\mathrm{Allen}} \}$. In this figure, we compare the weight parameters $w$ separately for the Senkov and Cantor alloy groups. The x-axis represents the weight parameter $w$, whereas the y-axis corresponds to the feature labels. The figure shows that the weighting trends differ between the two alloy groups. For the Senkov alloy system, a smaller variation in electronegativity appears to be important, whereas in the Cantor alloy system, a smaller variation in atomic size is emphasized.

In HEAs, the variation in elemental properties among constituent atoms affects phase stability. The Cantor alloy adopts a face-centered cubic (FCC) lattice, which requires dense atomic packing and lattice coherence. Therefore, minimizing atomic size mismatch is critical, as this helps to suppress misfit strain energy caused by lattice distortion and promotes the formation of a homogeneous solid solution. In the Hume-Rothery rules \cite{HumeRothery1934}, limiting the difference in atomic radii to within 15\% is considered a crucial condition for the formation of solid solutions. The compatibility of atomic sizes directly contributes to the stabilization of the crystal structure. By contrast, Senkov alloys primarily exhibit a body-centered cubic (BCC) lattice, which structurally allows for larger atomic size mismatches. However, large variations in electronegativity can lead to an inhomogeneous electron density distribution, potentially resulting in local chemical instability. Consequently, a system comprising elements with similar electronegativity tends to have stable metallic bonding, maintaining a single-phase structure even under high-temperature conditions.

These observations suggest that the effects of different types of property variation on phase stability in each alloy system are consistent with the respective structural and chemical requirements. The Senkov alloys benefit from chemical uniformity whereas the Cantor alloys require size uniformity for structural coherence. Thus, the identified trends in feature importance are physically and chemically reasonable.

\subsection{Visualization of the selected feature space}
\begin{figure}
    \centering
    \includegraphics[width=\linewidth]{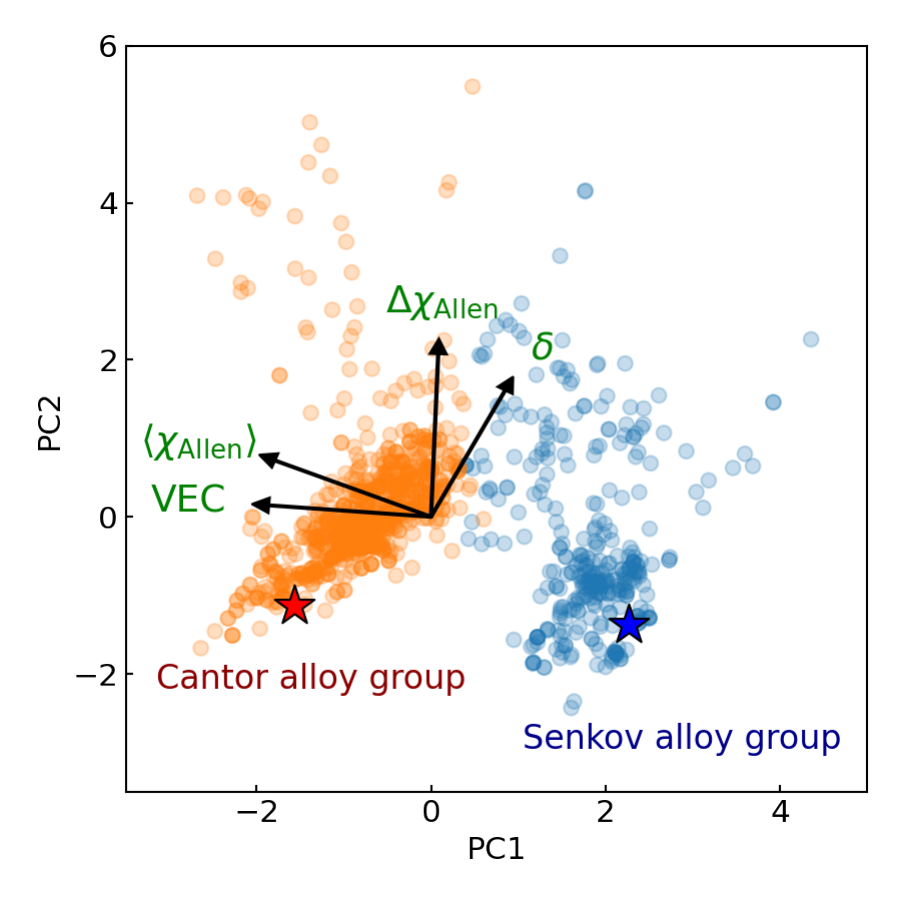}
    \caption{ Data visualization through dimensional reduction applying principal component analysis (PCA) to the optimal feature subset $\{ \delta, VEC, \langle \chi_{\mathrm{Allen}} \rangle, \Delta\chi_{\mathrm{Allen}} \}$. Black arrows are principal component loadings, indicating the contribution of each variable. Stars denote anchor data (of the Senkov alloy and Cantor alloy). }
    \label{fig:clustering}
\end{figure}

This section presents a visualization of the feature space identified by the best-performing model. Figure \ref{fig:clustering} shows the data projected via principal component analysis (PCA) applied to the optimal feature subset $\{ \delta, VEC, \langle \chi_{\mathrm{Allen}} \rangle, \Delta\chi_{\mathrm{Allen}} \}$. The stars denote the anchor data (of the Senkov alloy and Cantor alloy). The x- and y-axes show the first principal component (PC1) and second principal component (PC2), respectively. The black arrows represent principal component loadings, indicating the contribution of each variable to the principal components. 

The contribution rates of PC1 and PC2 were 47.34\% and 31.21\%, respectively, with the two components collectively accounting for 80\% of the variance. The colors represent clustering results based on the Euclidean distance from the anchor data within the four-dimensional feature space. In Figure \ref{fig:clustering}, it appears that the Senkov alloy group and Cantor alloy group are separated along the axes of the valence electron concentration $VEC$ and the average Allen electronegativity $\langle \chi_{\mathrm{Allen}} \rangle$. In contrast, $\delta$ seems to have a minimal effect on the group division. Moreover, Figure \ref{fig:coef} shows a small weight coefficient of $VEC$. This result and Figure \ref{fig:clustering} imply that $VEC$ is an important feature for group division rather than for single-phase and multi-phase prediction. The effectiveness of $VEC$ as a discriminative descriptor between the Senkov and Cantor alloys is well grounded in materials physics. $VEC$ affects the stability of crystal structures through its effect on the electronic structure-empirically, alloys with $VEC \gtrsim 8$ tend to stabilize in an FCC lattice whereas those with $VEC \lesssim 6.87$ favor a BCC lattice \cite{vec2011,vec2015}. The Cantor alloys typically exhibit an FCC lattice owing to their high $VEC$, whereas Senkov alloys often have BCC lattice or multiphase structures associated with lower $VEC$ values.

\section{Limitations and scope}
The proposed method depends on manually selected anchor points, which affect how data are clustered and how local models are formed. As a result, the performance may vary depending on the choice of anchors.

In addressing this issue, our Bayesian framework---particularly the use of the WBIC---can be extended to compare different anchor sets. In future work, we will systematically infer optimal anchor combinations by evaluating their Bayesian free energy, making the model more robust and data-driven.

\section{Conclusion}
We constructed a mixture linear model incorporating feature selection, with a focus on dataset structure characteristics based on a greedy search, using HEAs as a case study. The model demonstrates effective screening capability, using only four descriptors: $\{ \delta, \mathrm{VEC}, \langle \chi_{\mathrm{Allen}} \rangle, \Delta\chi_{\mathrm{Allen}} \}$. 

Furthermore, our method clearly indicated that different models were constructed for Cantor-type alloy and Senkov-type alloy, suggesting that the underlying factors driving high-entropy behavior vary between these two alloy families. Specifically, a smaller variance in electronegativity was found to be critical for Senkov-type alloys whereas a smaller atomic size mismatch was shown to be more influential for Cantor-type alloys. The proposed method is not limited to HEAs and can be readily applied to other materials datasets exhibiting exploration bias, such as those generated through iterative experimental design or data-driven screening.
% 本研究では，ハイエントロピー合金を題材に，「貪欲的探索に基づくデータセット構造の特性」に着目した特徴量選択を伴う混合線形モデルの構築を目的とした．結果として，4つの特徴量$\{ \delta, VEC, \langle \chi_{\mathrm{Allen}} \rangle, \Delta\chi_{\mathrm{Allen}} \}$のみで十分なスクリーニング効果を発揮するモデルの構築に成功した．さらに，本研究で得られたモデルでは，Cantor合金とSenkov合金のそれぞれに類似した合金で異なるモデルが構築されたことが確認された．この結果は，Cantor合金系とSenkov合金系でハイエントロピー合金になる因子が異なることを示唆している．具体的には，Senkov合金系では「電気陰性度のばらつきが小さい」ことが重要であり，Cantor合金系では「原子サイズ差のばらつきが小さい」ことが重要であることが示された．本研究の解析モデルは，ハイエントロピー合金だけでなく，他の材料データセットにも効果的であることが期待される．

\section*{Acknowledgments}
We thank Dr. Kenji Nagata (NIMS) and Dr. Syuki Yamanaka (Hokkaido University; currently at Bridgestone Corporation) for their insightful discussions, Mrs. Miya Kojima (NIMS) for her valuable assistance with data collection and organization.

\bibliographystyle{tfnlm} % 参考文献
\bibliography{main} %

\section{Appendix}

% \noindent\textbf{Appendix A. High entropy alloy dataset outline }\medskip
\subsection{Outline of the high-entropy alloy dataset}\label{appx:dataset}
\begin{figure}
    \centering
    \includegraphics[width=\linewidth]{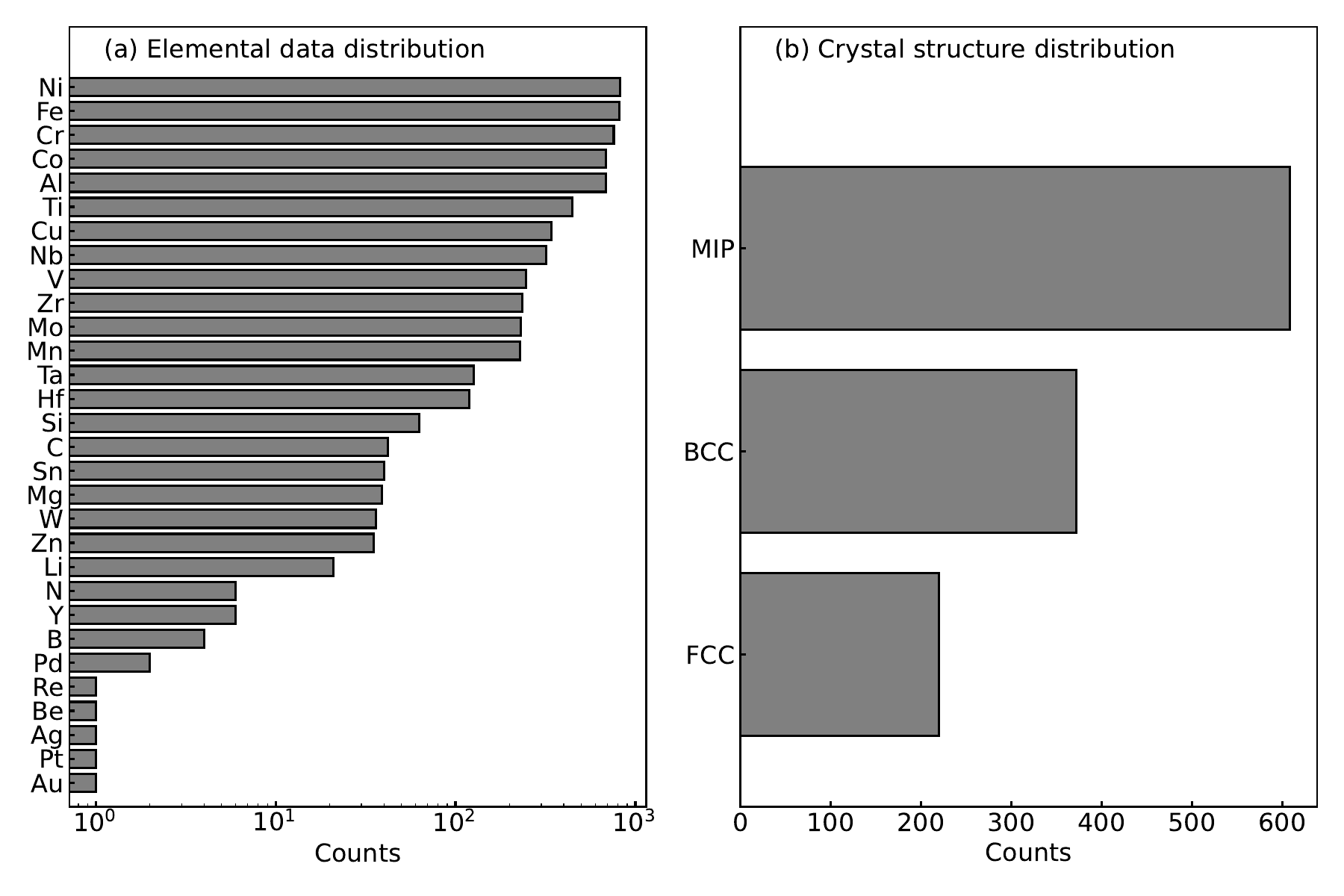}
    \caption{Data distribution of the dataset used in this study [(a) Elemental data distribution, (b) crystal structure distribution]. MIP refers to a mixture of intermetallic phases. The single phase has a BCC or FCC structure.}
    \label{fig:data}
\end{figure}

This section outlines the high-entropy alloy (HEA) dataset. We used open data \cite{Singh2023} in this study. Figure \ref{fig:data} shows the data distribution of the HEA dataset. Figures \ref{fig:data} (a) and (b) show the elemental data distribution and crystal structure distribution, respectively. In Figure \ref{fig:data} (b), MIP denotes the mixture of intermetallic phases. The MIP refers to a mixture of intermetallic phases.

As shown in Figure \ref{fig:data} (a), there is an imbalance in the elemental representation, with the number of occurrences per element ranging from just 1 to nearly 1000. This pronounced variation likely results from a biased sampling strategy, such as greedy searching, which tends to favor elements that are already known to form stable or interesting HEA phases.

Figure \ref{fig:data} (b) shows that the dataset contains a roughly balanced number of samples for multi-phase (MIP) and single-phase (BCC or FCC) structures. Among the single-phase entries, BCC lattices are more prevalent than FCC lattices, indicating a possible preference or stability tendency toward BCC lattices in the sampled compositions.

% \noindent\textbf{ Appendix B. Metrics based on our model }\medskip
\subsection{Metrics based on our model}\label{appx:metrics}
This section presents metrics based on our model. Our model is linear and low-dimensional (having four features). The predicted labels (single- or multi-phase) were obtained by thresholding the output probabilities $\bm{p}$ at 50\%. Table \ref{table:confusion_matrix} presents the confusion matrix based on our model whereas Table \ref{table:metrics} presents the representative metrics (precision, recall, and F1-score) of the classification score based on our model. Note that in binary classification, recall of the positive class is also known as "sensitivity" and recall of the negative class is "specificity." Support is the number of occurrences of each class in label $\bm{y}$. In the tables, the first score is the score for the test data whereas the score in parenthesis is the score for the training data.

\begin{table}
    \centering
    \caption{Confusion matrix based on our model. The first score is the score for the test data whereas the score in parenthesis is the score for the training data.}
    \label{table:confusion_matrix}
    \begin{tabular}{c|cc}
        & Predicted Positive & Predicted Negative \\
    \hline
    Actual Positive & 46 (440) & 9 (112) \\
    Actual Negative & 28 (210) & 37 (318) \\
    \hline
    \end{tabular}
\end{table}

\begin{table}
    \centering
    \caption{Metrics of the classification score based on our model. The first score is the score for the test data whereas the score in parenthesis is the score for the training data.}
    \label{table:metrics}
    \begin{tabular}{c|ccc|c}
    \hline
     & Precision & Recall & F1-score & Support \\
    \hline
    Multi-phase  & 0.62 (0.68) & 0.84 (0.80) & 0.71 (0.73) & 55 (553) \\
    Single-phase & 0.80 (0.74) & 0.57 (0.60) & 0.67 (0.66) & 65 (527) \\
    \hline
    Average      & 0.71 (0.71) & 0.70 (0.70) & 0.69 (0.70) & 120 (1080) \\
    \hline
    \end{tabular}
\end{table}

% \noindent\textbf{ Appendix C. Model selection using the marginal probability}\medskip\
\subsection{Model selection using the marginal probability}\label{appx:selection}
In this section, we compare three models---the fixed effects model, the random effects model, and the combined fixed and random effects model---based on their marginal probabilities in the context of high-entropy alloys. In this framework, fixed effects capture influences that are consistent across the entire dataset, whereas random effects account for variations specific to particular groups, reflecting inter-group differences within a mixed modeling approach. 

The differences between the three models are shown below. In the fixed effects model (the simple model in the main text), the prior distribution $P(\bm{w})$ is $\prod_{m=1}^{|\mathcal{S}|}{P(w_m)}$
\begin{align}\label{eq:fixed}
    w_{m} \sim P(w_{m}) = \mathcal{N}(w_{m} \mid \mu_{\mathrm{N}} = 0.0,\sigma^{2}_{\mathrm{N}} = 1.0).
\end{align}
The fixed effects model comprises a single linear model. In the random effects model (the mixed model in the main text), the prior distribution $P(\bm{w})$ is $\prod_{m=1}^{|\mathcal{S}|}{\prod_{k=1}^{K}{P(w_{mk})}}$:
\begin{align}\label{eq:random}
    w_{mk} \sim P(w_{mk}) = \mathcal{N}(w_{mk} \mid \mu_{\mathrm{N}} = 0.0,\sigma^{2}_{\mathrm{N}} = 1.0).
\end{align}
The fixed effects model learns a different weight coefficient for each group. In the fixed and random effects model, the prior distribution $P(\bm{w})$ and $P(\bm{\mu})$ are $\prod_{m=1}^{|\mathcal{S}|}{\prod_{k=1}^{K}{P(w_{mk})}}$ and $\prod_{m=1}^{|\mathcal{S}|}{P(\mu_m)}$:
\begin{equation}\label{eq:fixed_random}
\begin{aligned}
    w_{mk} &\sim P(w_{mk}) = \mathcal{N}(w_{mk} \mid \mu_{\mathrm{N}} = \mu_m,\sigma^{2}_{\mathrm{N}} = 0.1),\\
    \mu_{m} &\sim P(\mu_{m}) = \mathcal{N}(\mu_{m} \mid \mu_{\mathrm{N}} = 0.0,\sigma^{2}_{\mathrm{N}} = 1.0).
\end{aligned}
\end{equation}
The fixed and random effects model learns different weight coefficients for each group, however the average value of the prior distribution has a common value, leading to similar weight coefficients for each group. Here, the probability distribution $\mathcal{N}(\mu_{\mathrm{N}}, \sigma^{2}_{\mathrm{N}})$ is a normal distribution with mean $\mu_{\mathrm{N}}$ and standard deviation $\sigma^{2}_{\mathrm{N}}$. The models referred to in the main text correspond to the fixed effects model (\ref{eq:fixed}) and the random effects model (\ref{eq:random}).

Table \ref{table:mixed_selection} shows the marginal probability and the Bayesian free energy in the three models. It is seen that the mixed model (having fixed and random effects) is better than other models. Note that the marginal probability is the result of performing marginalization operations on parameters and feature subsets. In addition, marginalization operations on parameters were performed using the WBIC.

\begin{table}
    \centering
    \caption{ Bayesian free energy $F_{\mathrm{model}}$ and marginal probability $P_{\mathrm{model}}$ [\%] of three models (the fixed model, the random model and, the fixed and random model). The bias term $b$ is included in the parameters for the three models. }
    \label{table:mixed_selection}
    % \scalebox{1.1}[1.1]{
    \begin{tabular}{cccc|cc}
        \hline
        Eq. & Model & Effects & Parameter set & $F_{\mathrm{model}}$ & $P_{\mathrm{model}}$ [\%] \\
        \hline
        (\ref{eq:fixed}) & simple & fixed & $\bm{w} \in \mathbb{R}^{|\mathcal{S}|}$ & 737.72      & 0.00 \\
        (\ref{eq:random}) & mixed & random & $\bm{w} \in \mathbb{R}^{|\mathcal{S}|\times K}$ & \bf{685.82} & \bf{100.00} \\
        (\ref{eq:fixed_random}) & mixed & fixed \& random & $\bm{\mu} \in \mathbb{R}^{|\mathcal{S}|}, \bm{w} \in \mathbb{R}^{|\mathcal{S}|\times K}$ & 734.71      & 0.00 \\
        \hline
    \end{tabular}
    % }
    % \begin{tablenotes}
    %   \footnotesize
    %   \item \hfill*Our method
    % \end{tablenotes}
\end{table}

\subsection{Full vs selected feature space}
\begin{figure}
    \centering
    \includegraphics[width=\linewidth]{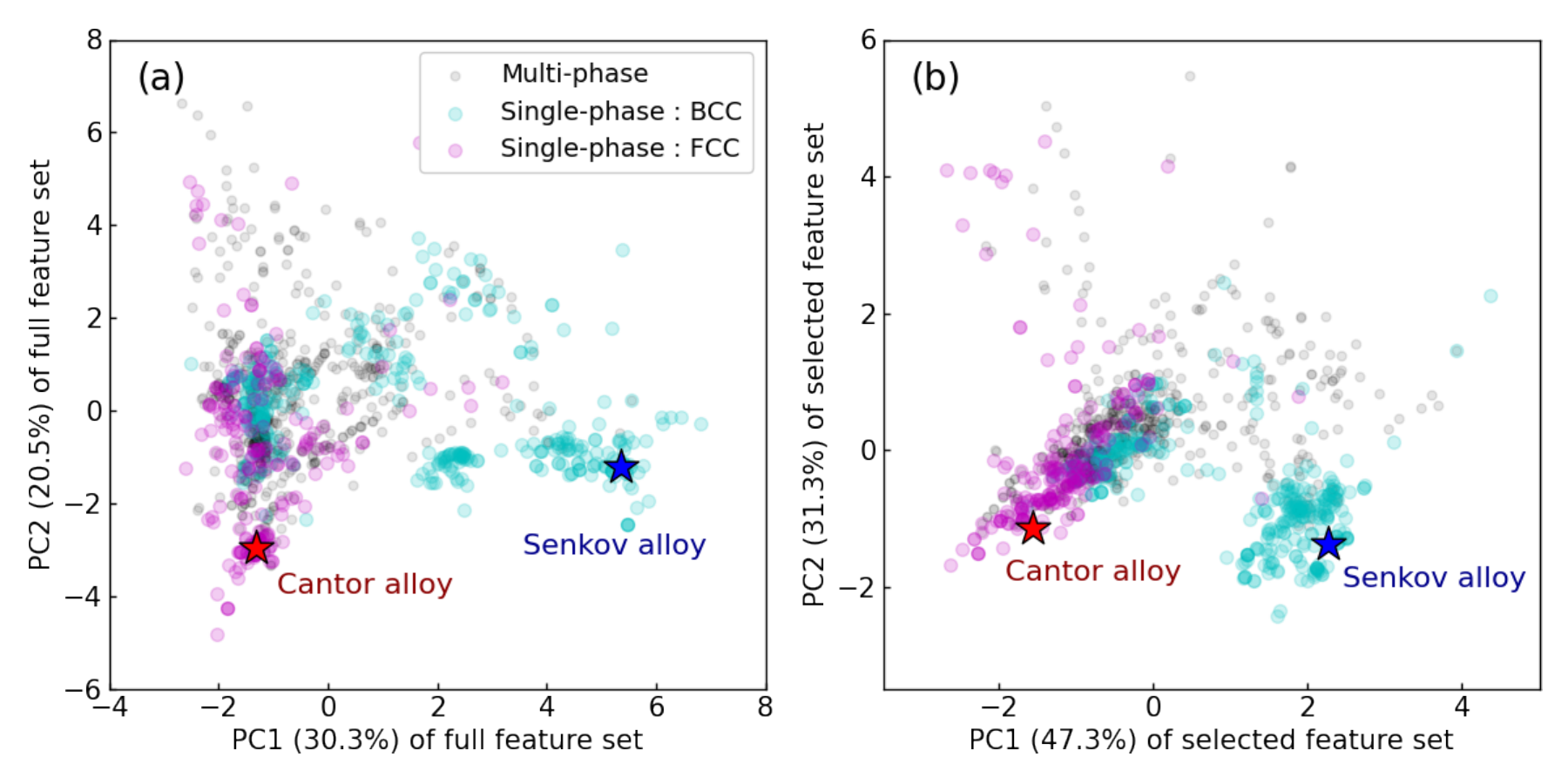}
    \caption{Data visualization via dimensionality reduction adopting principal component analysis (PCA) [(a) full feature set and (b) the selected optimal feature subset $\{ \delta, VEC, \langle \chi_{\mathrm{Allen}} \rangle, \Delta\chi_{\mathrm{Allen}} \}$]. Stars denote the anchor data (of the Senkov alloy and Cantor alloy).}
    \label{fig:compare_pca}
\end{figure}

This section presents visualizations of both full and selected feature spaces. Figure \ref{fig:compare_pca} shows dimensionality reduction results obtained through principal component analysis (PCA). Panels (a) and (b) present PCA results for the full feature set and selected optimal feature subset $\{ \delta, VEC, \langle \chi_{\mathrm{Allen}} \rangle, \Delta\chi_{\mathrm{Allen}} \}$, respectively. Stars denote the anchor data (of the Senkov alloy and Cantor alloy). The x- and y-axes give the first principal component (PC1) and second principal component (PC2), respectively. The percentages on the labels indicate the contribution rate of PCA. The selected feature space in (b) exhibits a more distinct cluster structure compared with the full feature space in (a).

\end{document}